\documentclass[amsmath,11pt]{article}

\usepackage{graphicx}    
\usepackage{amssymb}
\usepackage{amsfonts}
\usepackage{bm}
\usepackage{amsmath}

\newtheorem{theorem}{Theorem}[section]
\newtheorem{lemma}[theorem]{Lemma}

\newtheorem{remark}[theorem]{Remark}

\def\real {{\mathcal {I}}}
\def\rea {{I \!\! R}}

\def\Ci{P(|w|)}
\def\Cii{P(|w_*|)}
\def\Di{D(|w|)}
\def\Dii{D(|w_*|)}

\def\Ball{ {\mathcal {B}}}

\def\bigM{ {\mathcal {M}}}
\def\bigF{ {\mathcal {F}}}

\def\g{\hat g}

\def\Chi{\chi}
\newcommand{\rf}[1]{(\ref{#1})}

\def\g{\gamma}

\def\a{\alpha}

\def\s{\sigma}

\def\({\left(}
\def\){\right)}

\def\be#1\ee{\begin{equation}#1\end{equation}}
\def\ba#1#2\ea{\begin{array}#1#2\end{array}}
\def\bgr#1\egr{{\allowdisplaybreaks\begin{gather}#1\end{gather}}}
\def\bma#1\ema{{\allowdisplaybreaks\begin{align}#1\end{align}}}

\def\oplem#1{\begin{lemma}\, {\rm #1}\, \it }
\def\cllem{\end{lemma}\rm \par }
\def\opthm#1{\begin{theorem}\, {\rm #1}\, \it }
\def\clthm{\end{theorem}\rm \par }

\newcommand{\fer}[1]{(\ref{#1})}
\newcommand{\bq}{\begin{equation}}
\newcommand{\eq}{\end{equation}}
\def\bqa{\begin{eqnarray}}
\def\eqa{\end{eqnarray}}
\def\bd{\begin{displaymath}}
\def\ed{\end{displaymath}}

\begin{document}
\title{Kinetic models of opinion formation}

\author{
Giuseppe Toscani\thanks{Department of Mathematics, University of
Pavia, via Ferrata 1, 27100 Pavia, Italy.
\texttt{giuseppe.toscani@unipv.it}} }

\maketitle


\begin{abstract}
We introduce and discuss certain kinetic models of (continuous) opinion formation involving both exchange of
opinion between individual agents and diffusion of information. We show conditions which ensure that the kinetic
model reaches non trivial stationary states in case of lack of diffusion in correspondence of some opinion point.
Analytical results are then obtained by considering a suitable asymptotic limit of the model yielding a
Fokker-Planck equation for the distribution of opinion among individuals. Numerical results on the kinetic model
confirm the previous analysis.
\end{abstract}



\section{Introduction}

Microscopic models of both social and political phenomena
describing collective behaviors and self--organi\-za\-tion in a
society have been recently introduced and analyzed by several
authors \cite{GGS, Lig, Och, Sta, SO, SWS, Wei}.  The leading idea
is that collective behaviors of a society composed by a
sufficiently large number of individuals (agents) can be hopefully
described using the laws of statistical mechanics as it happens in
a physical system composed of many interacting particles. The
details of the social interactions between agents then
characterize the emerging statistical phenomena.

Among others, the modelling  of opinion formation attracted the
interest of a increasing number of researchers (cfr. \cite{DNAW,
Lig, Och, SWS} and the references therein). The starting point of
a large part of these models, however, is represented by a
cellular automata, where the lattice points are the agents, and
where any of the agents of a community is initially associated
with a random distribution of numbers, one of which is the
opinion. Hence society is modelled as a graph, where each agent
interacts with his neighborhoods in iterative way.

Very recently, other attempts have been successfully applied \cite{Ben, SL},
with the aim to describe formation of opinion by means of mean fields model
equations. These models are in general described by systems of ordinary
differential equations or partial differential equations of diffusive type,
that can in some case be treated analytically to give explicit steady states.
In \cite{Ben}, attention has been focused on two aspects of opinion formation,
which in principle could be responsible of the formation of coherent
structures. The first one is the remarkably simple compromise process, in which
pairs of agents reach a fair compromise after exchanging opinions \cite{BKR,
BKVR, DNAW, For1, HK, SM, WDAN}.

The second one is the diffusion process, which allows individual agents to
change their opinions in a random diffusive fashion. While the compromise
process has its basis on the human tendency to settle conflicts, diffusion
accounts for the possibility that people may change opinion through a global
access to information. In the present time, this aspect is gaining in
importance due to the emerging of new possibilities (among them electronic mail
and web navigation \cite{Ras}).

Following this line of thought,  we consider here a class of kinetic models of
opinion formation, based on two-body interactions involving both compromise and
diffusion properties in exchanges between individuals.  Compromise and
diffusion will be quantified by two parameters, which are mainly responsible of
the behavior of the model, and allow for a rigorous asymptotic analysis. In
consequence of our assumptions on the microscopic interaction, in various
relevant cases the model will satisfy mass and momentum conservation, which are
the starting point for studying the asymptotic behavior.

In this direction, we shall show that the kinetic model gives in a
suitable asymptotic limit (hereafter called quasi-invariant
opinion limit) a partial differential equation of Fokker-Planck
type for the distribution of opinion among individuals. Similar
diffusion equations were  obtained recently in \cite{SL} as the
mean field limit of the Ochrombel simplification of the Sznajd
model \cite{SWS}.

The equilibrium state of the Fokker-Planck equation can be
computed explicitly and reveals formation of picks in
correspondence to the points where diffusion is missing.

The mathematical methods we use are close to those used in the
context of kinetic theory of granular gases, where the limit
procedure is known as quasi-elastic asymptotics \cite{McNY2, to1}.
We mention here that a similar asymptotic analysis was performed
on a kinetic model of a simple market economy with a constant
growth mechanism \cite{CPT, Sl}, showing formation of steady
states with Pareto tails \cite{Pa}. In this context, the mean
field approximation leads to the same Fokker-Planck equation
\cite{BM, So}, showing consistency between  kinetic and stochastic
approaches.

The paper is organized as follows. In the next section we
introduce the binary interaction between agents, which is at the
basis of the kinetic model. The main properties of the model are
discussed in section 3. These properties justify the
quasi-invariant opinion limit procedure, performed in section 4.
The limit is illustrated by several choices of the diffusion  in
section 5.


\section{Description of the kinetic model}\label{model}

The goal of the forthcoming kinetic model of opinion formation, is to describe the evolution of the distribution of
opinions in a society by means of {\em microscopic} interactions among agents or individuals which exchange
information. To fix ideas, we associate opinion with a variable $w$ which varies continuously from $-1$ to $1$,
where $-1$ and $1$ clearly denote the two (extreme) opposite opinions. We will moreover assume that interactions do
not destroy the bounds,  which corresponds to impose that the extreme opinions can not be crossed. This crucial
rule emphasizes the difference between the present {\em social} interactions, where not all outcomes are permitted,
and the classical interactions between molecules, familiar to people working in kinetic theory of rarefied gases
\cite{Cer}.

Let $\real = [-1, +1]$ denote the interval of possible opinions.
From a microscopic view point, we describe the binary interaction
by the rules
 \bqa \nonumber
w'  & =  & w - \gamma\Ci( w- w_*) + \eta \Di \\
\label{trade_rule}
\\[-.25cm]
 w_*'  & =&  w_* - \gamma \Cii(w_*- w) + \eta_* \Dii \nonumber
 \eqa
 where the pair $(w,w_*)$, with $w, w_* \in \real$
denotes the opinions of two arbitrary individuals before the
interaction and $(w',w_*')$ their opinions after exchanging
information between them and with the exterior. In
(\ref{trade_rule}) we will not allow opinions to cross boundaries,
and thus the interaction takes place only if both $w', w'_* \in
\real$. In (\ref{trade_rule}) the coefficient $\gamma\in (0,1/2)$
is a given constant, while $\eta$ and $\eta_*$ are random
variables with the same distribution with variance $\sigma^2$ and
zero mean, taking values on a set $\Ball \subseteq \rea$.  The
constant $\g$ and the  variance $\sigma^2$ measure respectively
the compromise propensity and the modification of opinion due to
diffusion. Finally, the functions $P(\cdot)$ and $D(\cdot)$
describe the local relevance of the compromise and diffusion for a
given opinion.

Let us describe the details of the interaction in the right hand side. The
first part is related to the compromise propensity of the agents,  and the last
contains the diffusion effects of external events. Note that the
pre-interaction opinion $w$ increases (getting closer to $w_*$) when $w_* >w$
and decreases in the opposite situation. The presence of both the functions
$P(\cdot)$ and $D(\cdot)$ is linked to the hypothesis that the availability to
the change of opinion is linked to the opinion itself, and decreases as soon as
one gets closer to extremal opinions. This corresponds to the realistic idea
that extremal opinions are more difficult to change. We will present later on
various realizations of these functions. In all cases, however, we assume that
both $\Ci$ and $\Di$ are non increasing with respect to $|w|$, and in addition
$0 \le \Ci \le 1$, $0 \le \Di \le 1$.

In absence of the diffusion contribution ($\eta, \eta_*\equiv 0$), \fer{trade_rule} implies
  \bqa \nonumber
 w' + w'_* &=& w + w_*  + \g (w-w_*) \left( \Ci - \Cii \right)\\
\label{tr+-}
\\[-.25cm]
\nonumber w'-w'_* &=& \left(1- 2\g( \Ci + \Cii)\right)(w-w_*).
 \eqa
Thus, unless the function $P(\cdot)$ is assumed constant, $P=1$,
the total {\em momentum} is not conserved and it can increase or
decrease depending on the opinions before the interaction. If
$P(\cdot)$ is assumed constant, the conservation law is
reminiscent of analogous conservations which take place in kinetic
theory. In such a situation, thanks to the bounds on the
coefficient $\g$, equations (\ref{trade_rule}) correspond to a
granular gas like interaction (or to a traffic flow model
\cite{KW}) where the stationary state is a Dirac delta centered in
the average opinion (usually referred to as synchronized traffic
state in traffic flow modelling). This behavior is a consequence
of the fact that, in a single interaction, the compromise
propensity implies that the difference of opinion is diminishing,
with $|w'-w'_*| = (1-2\gamma)|w-w_*|$. Thus all agents will end up
in the society with exactly the same opinion. Note that in this
elementary case a constant part of the relative opinion is
restituted after the interaction. This property does not remain
true if the function $P$ depends on the opinion variable. In this
case
 \[
|w'-w'_* |= \left(1- 2\g( \Ci + \Cii)\right)|w-w_*|.
 \]
In fact, since $\gamma\in (0,1/2)$ and $0 \le \Ci \le 1$,  $0 \le \Di \le 1$,
 \[
0 \le \varepsilon( w,w_*) = 1- 2\g( \Ci + \Cii) \le 1.
 \]
Hence the general case corresponds to a granular gas interaction with a variable coefficient of restitution
\cite{to1}.

 We remark moreover that, in absence of diffusion, the lateral bounds  are not violated, since
 \bqa \nonumber
w'  & =  & (1 - \gamma\Ci) w + \gamma\Ci w_* \\
\label{gran-rule}
\\[-.25cm]
\nonumber w_*'  & =&   (1 - \gamma\Cii ) w_* + \gamma\Cii w
 \eqa
 implies
 \[
 \max\left\{|w'|, |w'_*|\right\} \le  \max\left\{|w|,
 |w_*|\right\}.
 \]

Let  $f(w,t)$ denote the distribution of opinion $w \in \real$ at time $t \ge 0$. A direct application of standard
methods of kinetic theory of binary interactions \cite{Cer} allows to recover the time evolution of $f$ as a
balance between bilinear gain and loss of opinion terms, described by the integro-differential equation of
Boltzmann type
 \be \frac{\partial f}{\partial t} =
\int_{\Ball^2}\int_\real \left('\beta \frac 1J f('w)
 f('w_*) - \beta f(w) f(w_*)\right) d
 w_*\,d\eta\,d\eta_* ,\label{eq:boltz}
  \ee
where $('w,'w_*)$ are the pre-interaction opinions that generate the couple
$(w,w_*)$ of opinions after the interaction. In (\ref{eq:boltz}) $J$ is the
Jacobian of the transformation of $(w,w_*)$ into $(w',w'_*)$,
  while the kernels $'\beta$  and
$\beta$ are related to the details of the binary interaction.

As usual in classical kinetic theory of rarefied gases, the
interaction integral on right-hand side of \fer{eq:boltz}
represents  the instantaneous variation of the distribution of
opinion, due to the binary exchanges of information. The presence
of the Jacobian $J$, guarantees that equation \fer{eq:boltz}
preserves the mass (total opinion), for any choice of the rate
function $\beta$.  The transition rate is taken of the form
 \be\label{kernel}
\beta_{(w,w_*) \to (w',w_*')}=\Theta(\eta)\Theta(\eta_*)\chi(|w'|
\le 1)\chi(|w'_*| \le 1),
 \ee
where $\chi(A)$ is the indicator function of the set $A$, and $\Theta(\cdot)$ is a symmetric probability density
with zero mean and  variance $\sigma^2$. The rate function $\beta_{(w,w_*) \to (w',w_*')}$ characterizes the
effects of external events on opinion through the distribution of the random variables $\Theta$ and $\Theta_*$ and
takes into account the hypothesis that bounds can not be violated. We remark that in principle the support $\Ball$
of the symmetric random variable is a subset of $\real$, to prevent diffusion to generate a complete change of
opinion. This property can  be weakened by assuming for example diffusion as a random variable normally distributed
but well concentrated on zero.

For a general probability density $\Theta(\cdot)$, the rate function $\beta$ depends on the opinion variables
$(w,w_*)$ through the indicator functions $\Chi$. This fact reminds a similar  property of the classical Boltzmann
equation \cite{Cer, DPR}, where the rate function depends on the relative velocity. As we shall see, a simplified
situation occurs when a suitable choice of the function $D(\cdot)$ in \fer{trade_rule} coupled with a small support
$\Ball$ of random variables implies that both $|w'| \le 1$ and $|w'_*| \le 1$, and the kernel $\beta$ does not
depend on the opinion variables $(w,w_*)$. In this case the kinetic equation (\ref{eq:boltz}) is the corresponding
of the classical Boltzmann equation for Maxwell molecules \cite{Bob}, which presents several mathematical
simplifications. In all cases however, methods borrowed from kinetic theory of rarefied gas can be used to study
the evolution of the function $f$.


\section{Simplifications and main properties of the model}
The main problem in opinion dynamics is the formation of stationary profiles
for the opinion. In the kinetic picture this corresponds to an investigation of
the large time behavior of the density of opinion $f(w,t)$. To investigate in
detail the large-time behavior, a preliminary analysis of equation
\fer{eq:boltz} is needed. We will start this analysis by introducing some
notations and by discussing the main properties of the kinetic equation.

 Let $Q(f,f)$ denote the interaction integral,
 \be
 Q(f,f)(w) = \int_{\Ball^2}\int_\real \left('\beta \frac 1J f('w)
 f('w_*) - \beta f(w) f(w_*)\right) d
 w_*\,d\eta\,d\eta_*  .
  \ee

 Let $\bigM_0(A)$ the space of all probability measures taking values in $A \subseteq\rea$ and by
  \bq\label{misure} \bigM_{p}(A) =\left\{
\Theta \in\bigM_0: \int_{A} |w|^{p}d\Theta(w) < +\infty, p\ge
0\right\},
 \eq
 the space of all Borel
probability measures of finite momentum of order $p$, equipped with the topology of the weak convergence of the
measures.

Let $\bigF_s(\real)$,  be the class of all real  functions $h$ on
$\real$ such that $h(\pm 1)= h'(\pm 1)=0$, and  $h^{(m)}(v)$ is
H\"older continuous of order $\delta$,
 \bq\label{lip} \|h^{(m)}\|_\delta= \sup_{v\not= w} \frac{|h^{(m)}(v) -h^{(m)}(w)|}{
|v-w|^\delta} <\infty,
 \eq
 the integer $m$ and the number $0 <\delta \le 1$ are such that $m+\delta =s$, and
$h^{(m)}$ denotes the $m$-th derivative of $h$.

 In the rest of the paper we will assume that the symmetric probability density $\Theta(\eta)$
 which characterizes the diffusion of information  belongs to $\bigM_{2+\alpha}$, for some $\alpha
 >0$.  Moreover, to simplify  computations, we assume that this density is obtained
 from a given random variable $Y$ with zero mean and unit
 variance, that belongs to $\bigM_{2+\alpha}$. Thus, $\Theta$ of variance $\sigma^2$ is the density of
 ${\sigma}Y$. By this assumption, we can easily obtain the dependence on $\sigma$ of the moments of $\Theta$.
 In fact, for any $p >0$ such that the  $p$-th moment of $Y$ exists,
 \[
\int_{\rea}|\eta|^{p}\Theta(\eta) d\eta = E\left(
\left|{\sigma}Y\right|^{p}\right) =
\sigma^{p}E\left(\left|Y\right|^{p}\right).
\]
 By a weak solution of the initial value problem for equation \rf{eq:boltz},
corresponding to the initial probability density $f_0(w) \in
\bigM_{0}(\real)$, we shall mean any probability density $f \in
C^1(\rea_+, \bigM_{0}(\real))$ satisfying the weak form of the
equation
 \begin{eqnarray}\label{weak boltz}
&&\frac d{dt}\int_{\real} \phi(w)f(w,t)\,dw  = (Q(f,f),\phi) = \nonumber \\
 &&\int_{\real^2} \int_{\Ball^2} \beta_{(w,w_*) \to (w',w'_*)} f(w)
f(w_*) ( \phi(w')-\phi(w)) d w_* d w d \eta\,d\eta_*,
\end{eqnarray}
for $t>0$ and all $\phi \in \bigF_{p}(\real)$, and such that for
all $\phi \in \bigF_{p}(\real)$
 \bq\label{ic} \lim_{t\to 0} \int_{\real} \phi(w)f(w,t)\, dw = \int_{\real} \phi(w)f_0(w)\, dw.
  \eq
The form \fer{weak boltz} is easier to handle, and it is the starting point to
explore the evolution of macroscopic quantities (moments). By symmetry reasons,
we can alternatively use the symmetric form
 \begin{eqnarray} \label{weak boltz2}\nonumber
\frac{d}{dt} \int_\real f(w) \phi(w)\,dw &=& \frac12
\int_{\real^2} \int_{\Ball^2}
\beta_{(w,w_*) \to (w',w'_*)}  f(w) f(w_*) \\
\\[-.25cm]
\nonumber &&( \phi(w')+\phi(w'_*)-\phi(w)-\phi(w_*)) dw_*\, dw\,
d\eta\,d\eta_*.
  \end{eqnarray}
Existence of a weak solution to the initial value problem for equation
(\ref{eq:boltz}) can be easily obtained by using methods first applied to the
Boltzmann equation \cite{Cer}. On the other hand, for a general kernel
\fer{kernel},  it appears extremely difficult to describe in detail the
large-time behavior of the solution.

For this reason, at first we restrict our analysis to the cases in
which the kernel $\beta$ does not depend on the opinion variables
(Maxwellian case). In this direction, let us briefly discuss  the
importance of the support $\Ball$ of the probability density
function $\Theta(\eta)$ in connection with the possible
simplification of the kernel. To clarify the point, let us set
 \[
\Di = 1 -|w|.
 \]
 This function satisfies all the requirements we fixed
 in the previous Section.
 Then, since
  \[
  (1-\g\Ci)w +\g\Ci w_* + (1 -|w|)\eta \le (1-\g\Ci )w +\g\Ci  + (1 -|w|)\eta,
  \]
  in order that $|w'| \le 1$ it suffices that
  \be
(1-\g\Ci)w +\g\Ci  + (1 -|w|)\eta \le 1,
 \ee
 or, what is the same
  \be\label{cc}
 (1 -|w|)\eta \le (1-\g\Ci)(1-w).
 \ee
 Since $\Ci \le 1$,   bound \fer{cc} is verified for all $w \ge 0$, as soon as $\eta \le 1-\g$.
 Analogous result holds if $w \le 0$. Hence, if $ \Di = 1 -|w|$ and $\Ball = (-(1-\g),
 1-\g)$, both $w'$ and $w'_*$ belong to the right interval.

 \begin{remark}\label{max} {\rm  Any choice of $\Di$ and $\Ball$ which are suitable to preserve the lateral bounds of
extreme opinions allows to study in detail the dynamics of the model with a
significant simplification. In these cases, the kernel $\beta$ defined in
\fer{kernel} simplifies to
 \[
\beta_{(w,w_*) \to (w',w_*')}= \beta(\eta, \eta_*) =
\Theta(\eta)\Theta(\eta_*).
 \]
In the rest of the paper we will limit ourselves to such type of kernels. As
briefly explained in the previous Section, this assumption is the analogue to
Maxwell molecules interaction in the Boltzmann equation  \cite{Cer}.}
 \end{remark}

From (\ref{weak boltz}) (or equivalently from (\ref{weak boltz2})) conservation of the total opinion is obtained
for $\phi(w)=1$, which represents in general the only conservation property satisfied by the system.  The choice
$\phi(w)= w$ is of particular interest since it gives the time evolution of the average opinion. We have
 \be
\frac d{dt}\int_{\real} w f(w,t)\,dw  =
 \int_{\real^2} \int_{\Ball^2} \beta(\eta, \eta_*) f(w)
f(w_*) \g (\Ci w_* -\Ci w)d w_* d w d \eta\,d\eta_*  \nonumber
 \ee
 \[
 + \int_{\real^2} \int_{\Ball^2} \beta(\eta, \eta_*) f(w) f(w_*)  \eta \Di  d w_* d w d \eta\,d\eta_*
 \]
The first integral on the right--hand side represents the contribution of the exchange of information to the
variation of momentum. In case $\Ci = 1$, this contribution disappears, since, by symmetry,
  \[
\int_{\real^2} \int_{\Ball^2} \beta(\eta, \eta_*) f(w) f(w_*) \g
(w_* -w) d w_* d w d \eta\,d\eta_* =0.
  \]
In this case
 \[
 \frac d{dt}\int_{\real} w f(w,t)\,dw =
 \]
  \be\label{grow}
= \int_{\real^2} \int_{\Ball^2}\eta \Theta(\eta)\Theta(\eta_*)
\Chi(|w'|\leq 1)\Chi(|w'_*|\leq 1)\Di f(w) f(w_*) d w_* d w d
\eta\,d\eta_* = 0.
 \ee
 since the mean value of the density $\Theta$ is zero. This shows
that $P$ constant implies that the average opinion is conserved.
The situation changes when $P$ is not constant. In this case, the
 time evolution of the average opinion is given by
   \be\label{grow5}
  \frac d{dt}\int_{\real} w f(w,t)\,dw =
\g \int_{\real} \Ci f(w)\, d w \int_{\real} w f(w)\, d w -
\g\int_{\real} w\Ci f(w)\, d w.
 \ee
Note that equation \fer{grow5} is not closed.

 Let us fix now $\phi(w) = w^2$. We have
 \be
\frac d{dt}\int_{\real} w^2 f(w,t)\,dw  =
  \frac 12 \int_{\real^2} \int_{\Ball^2} \Theta(\eta)\Theta(\eta_*) f(w)
f(w_*) ({w'}^2 + {w'_*}^2 -w^2 -w_*^2) d w_* d w d \eta\,d\eta_*
 \ee
Taking in mind that $\Theta$ has zero mean and variance
$\sigma^2$, by easy computations one shows that
 \be
 \frac 12 \int_{\real^2} \int_{\Ball^2} \Theta(\eta)\Theta(\eta_*) f(w)
f(w_*) ({w'}^2 + {w'_*}^2 -w^2 -w_*^2 ) d w_* d w d \eta\,d\eta_*
= \nonumber
 \ee
 \be
  \g^2 \int_{\real^2} \Ci^2(w-w_*)^2 f(w)
f(w_*) \, d w d w_* -\nonumber
 \ee
 \be
 2\g\int_{\real^2} \Ci w(w-w_*) f(w) f(w_*) \,
d w d w_* + \sigma^2\int_{\real}\Di^2  f(w) d w.
 \ee
The choice $\Ci =1$ leads to the simpler evolution equation
 \be
\frac d{dt}\int_{\real} w^2 f(w,t)\,dw  = -2\g(1-\g)\left[
\int_{\real} w^2 f(w)\, d w - m^2 \right] +
\sigma^2\int_{\real}\Di^2  f(w) d w,
 \ee
where $m$ is the constant value of the average opinion
 \[
 m=  \int_{\real}w f(w,t) d w.
 \]
Since $|w| \le 1$, the boundedness of the mass implies that all
moments are bounded. This implies that, in all cases, we can draw
conclusions on the large--time convergence of the class of
probability densities $\left\{ f(w, t)\right\}_{t \ge 0}$. By
virtue of Prokhorov theorem (cfr. \cite{LR}) the existence of a
uniform bound on moments implies in fact that this class is tight,
so that any sequence $\left\{ f(w, t_n)\right\}_{n \ge 0}$
contains an infinite subsequence which converges weakly to some
probability measure $f_{\infty}$.

\section{The quasi-invariant opinion limit}

The analysis of the previous Section shows that in general it is quite
difficult both to study in detail the evolution of the opinion density, and to
describe its asymptotic behavior. For a general kernel one has in addition to
take into account that the mean opinion is varying in time.  As is usual in
kinetic theory, however, particular asymptotics of the equation result in
simplified models (generally of Fokker-Planck type), for which it is relatively
easier to find steady states, and to prove their stability. These asymptotics
are particularly relevant in case they are able to describe with a good
approximation the stationary profiles of the kinetic equation. In order to give
a physical basis to these asymptotics, let us discuss  the interaction rule
(\ref{trade_rule}) from a slightly different point of view. For the moment, we
will assume $\Ci =1$,  so that conservation both of mass and momentum holds.
The case of a general $\Ci$ will be treated subsequently. Let us denote by
$E(X)$ the mathematical expectation of the random variable $X$. Then the
following properties follow from (\ref{trade_rule})
 \be\label{mean}
 E[w' + w'_*] = w+w_*, \quad   \quad E[w' - w'_*] = (1-2\gamma)(w-w_*).
 \ee
The first equality in \rf{mean} describes the property of mean
conservation of opinion. The second refers to the compromise
propensity, which plays in favor of the decrease (in mean) of the
distance of opinions after the interaction. This tendency is a
universal consequence of the rule (\ref{trade_rule}), in that it
holds whatever distribution one assigns to $\Theta$, namely to the
random variable which accounts for the effects of the external
word in opinion formation.

The second property in \rf{mean} is analogous to the similar one
that holds in a collision between molecules in a granular gas.
There the quantity $e=2\gamma$ is called "coefficient of
restitution", and describes the peculiar fact that energy is
dissipated \cite{to1}.

We consider now the situation in which most of the interactions
produce a very small exchange of opinion ($\gamma \to 0$), while
at the same time  both properties \rf{mean} remain at a
macroscopic level. This corresponds to pretend that, while $\g \to
0$,
 \be\label{11}
\int_{\real^2}(w+w_*)f(w)f(w_*) dw dw_* = 2\int_{\real} w f(w) dw
= 2m(t)
 \ee
 remains constant, and
 \be\label{22}
\frac 12\int_{\real^2}(w-w_*)^2f(w)f(w_*) dw dw_* = \int_{\real}
w^2 f(w)\, d w - m_0^2 = C_f(t)
 \ee
 varies with time, and decays to zero when the diffusion is not present (i.e. $\sigma=0$).

Since in our case the kernel $\beta$ does not depend on the
opinion variables, \rf{grow} implies that $m(t)=m_0$ independently
of the value of $\g$. Moreover, using the computations of the
previous Section,  one obtains that $C_f(t)$ varies with law
 \be\label{33}
 \frac {dC_f(t)}{dt} =
 -2\g(1-\g) C_f(t) + \sigma^2\int_{\real}\Di^2 f(w) d w.
 \ee
Hence, if we set
 \be\label{resc}
 \tau = \gamma t, \quad g(w,\tau) = f(w,t),
 \ee
 which implies $f_0(w) = g_0(w)$, it follows
 \be
\label{45}
 \frac {dC_g(\tau)}{d\tau} = -2\left( 1-\gamma\right) C_g(\tau) +
 \frac{\sigma^2}\gamma \int_{\real}\Di^2  f(w) d w .
 \ee
 Letting now both $\gamma \to 0$ and $\sigma \to 0$ in such a way that $\sigma^2/\gamma =
 \lambda$, \rf{45} becomes in the limit
 \be
\label{46}
 \frac {dC_g(\tau)}{d\tau} = -2 C_g(\tau) +
 \lambda \int_{\real}\Di^2  f(w) d w .
 \ee
This  argument shows that the value of the ratio $\sigma^2/\gamma$
is of paramount importance to get asymptotics which maintain
memory of the microscopic interactions. It is remarkable that,
thanks to \fer{resc}, $t = \tau/\g$, so that the limit $\g \to 0$
describes the large-time behavior of $f(v,t)$. On the other hand,
since $f(w,t) = g(w,\tau)$ the large-time behavior of $f(w,t)$ is
close to the large-time behavior of $g(w, \tau)$.

\begin{remark} {\rm The balance $\gamma \to 0$ and $\sigma \to 0$ in such a way that $\sigma^2/\gamma =
 \lambda$, allows to recover in the limit the contributions
 due both to compromise propensity and the diffusion. Other limits
 can be considered, which are diffusion dominated ($\sigma^2/\gamma =
 \infty$) or compromise dominated ($\sigma^2/\gamma =
0$). As we shall present in the next section, however, the
formation of an asymptotic profile for the opinion is linked to
the first balance \cite{Ben}. }
\end{remark}

In the remainder of this section, we shall present the rigorous derivation of a
Fokker-Planck model, starting from the Boltzmann equation for the opinion
density $g(w, \tau)$, when both $\gamma \to 0$ and $\sigma \to 0$ in such a way
that $\sigma^2/\gamma \to \lambda $. For the sake of simplicity, we will assume
that $\Ci=1$.  This type of analysis is close to the one described in
\cite{CPT} for a kinetic model of wealth distribution in a open economy.

The scaled density  $g(v,\tau)=f(v,t)$ satisfies the equation (in
weak form)
 \be\label{evol}
 \frac{d}{d\tau} \int_\real g(w) \phi(w)\,dw
= \frac1{\gamma} \int_{\real^2} \int_{\Ball^2}
\Theta(\eta)\Theta(\eta_*) g(w) g(w_*) ( \phi(w')-\phi(w)) d w_* d
w d \eta\,d\eta_*.
 \ee
Given $0 <\delta \le \alpha$, let us set $\phi \in
\bigF_{2+\delta}(\real)$.

By (\ref{trade_rule}),
$$
w'   - w =   \gamma (w_* - w) + \eta \Di .
$$
Then, if we use a second order Taylor expansion of $\phi$ around $w$
 $$
\phi(w') - \phi(w) = (\gamma (w_* - w) + \eta \Di) \phi'(w) + {1
\over 2} \left(\gamma (w_* - w) + \eta \Di\right)^2 \phi''(\tilde
w),
 $$
where, for some $0 \le \theta \le 1$
 \[
 \tilde w = \theta w' +(1 -\theta)w .
  \]

 Inserting this expansion in the collision operator,  we get
 \be
\frac{d}{d\tau}\int_\real g(w) \phi(w)\,dw  =
\frac1{\gamma}\int_{\real^2}\int_{\Ball^2}\Theta(\eta)\Theta(\eta_*)
 [  (\gamma (w_* - w) + \eta \Di) \phi'(w)\nonumber +
  \ee
  \be\label{form}
  + {1 \over 2}
(\gamma (w_* - w) + \eta \Di)^2 \phi''(w) ]
 g(w)g(w_*)  d w_*\,d w\,d \eta\,d\eta_*  + R(\gamma, \sigma),
 \ee
where
 \begin{eqnarray*}
 R(\gamma, \sigma) &=
&\frac
1{2\gamma}\int_{\real^2}\int_{\Ball^2}\Theta(\eta)\Theta(\eta_*)
(\gamma (w_* - w) + \eta \Di)^2\cdot \\&& \cdot \left(
\phi''(\tilde w)- \phi''( w)\right)
 g(w)g(w_*)  d w_*\,d w\,d \eta\,d\eta_*
 \end{eqnarray*}
Since $\phi \in \bigF_{2+\delta}(\real)$, and $|\tilde w - w |=
\theta|w'-w|$
 \be\label{rem}
 \left| \phi''(\tilde w)- \phi''( w)\right| \le \| \phi''\|_\delta |\tilde w - w |^\delta \le
 \| \phi''\|_\delta |w' - w |^\delta .
  \ee
  Hence
 \begin{eqnarray*}
|R(\gamma, \sigma)| &\le& \frac{\|
\phi''\|_\delta}{2\gamma}\int_{\real^2}\int_{\Ball^2}\Theta(\eta)\Theta(\eta_*)\cdot
\\&& \cdot |\gamma (w_* - w) + \eta\Di |^{2+\delta} g(w) g(w_*) d
w_*\,d w\,d \eta\,d\eta_*
 \end{eqnarray*}
By virtue of the inequality
 \[
 |\gamma (w_* - w) + \eta \Di|^{2+\delta} \le 2^{1+\delta}
 \left( |\gamma( w_*- w)|^{2+\delta}+| \eta \Di|^{2+\delta}\right) \le
 \]
 \[
 \le 2^{2+\delta}\g^{1+\delta} + 2^{1+\delta}|\eta |^{2+\delta},
  \]
we finally obtain the bound
 \be\label{to0}
|R(\gamma, \sigma)| \le 2^{1+\delta}{\| \phi''\|_\delta}\left(
\gamma^{1+\delta} + \frac 1{2\gamma}
\int_{\Ball}|\eta|^{2+\delta}\Theta(\eta) d\eta \right)
 \ee
 Since $\Theta$ is a probability density with zero mean and
$\lambda\gamma$ variance, and $\Theta$ belongs to $\bigM_{2+\alpha}$, for  $\alpha
 >\delta$,
 \[
\int_{\real}|\eta|^{2+\delta}\Theta(\eta) d\eta = E\left( \left|\sqrt{\lambda\gamma}Y\right|^{2+\delta}\right) =
(\lambda\gamma)^{1+\delta/2}E\left(\left|Y\right|^{2+\delta}\right),
\]
and $E\left(\left|Y\right|^{2+\delta}\right)$ is bounded. Using
this equality into \rf{to0} one shows that $R(\gamma, \sigma)$
converges to zero as as both $\gamma$ and $\sigma$ converge to
zero, in such a way that $\sigma^2 = \lambda\gamma$. Within the
same scaling,
 \be
 \lim_{\gamma \to 0}
\frac1{\gamma}\int_{\real^2}\int_{\Ball^2}\Theta(\eta)\Theta(\eta_*)
 [  (\gamma (w_* - w) + \eta \Di) \phi'(w) + \nonumber
 \ee
 \be
 + {1 \over 2}
(\gamma (w_* - w) + \eta \Di)^2 \phi''(w) ]
 g(w)g(w_*)  d w_*\,d w\,d \eta\,d\eta_* = \nonumber
  \ee
  \be\label{FP1}
\int_{\real}\left[
 (m - w)  \phi'(w)  +
   \frac\lambda{2}\Di^2 \phi''(w)\right]
 g(w) d w
  \ee
Considering that $\phi \in \bigF_s(\real)$, we can integrate back
by parts. This shows that the right-hand side of \rf{FP1}
coincides with the weak form of the Fokker-Planck equation
 \be\label{FP}
 \frac{\partial g}{\partial \tau} = \frac \lambda 2\frac{\partial^2 }{\partial w^2}\left(\Di^2
 g\right) + \frac{\partial }{\partial w}\left((w -m)
 g\right).
 \ee
 Last, since the solution to the kinetic model conserves mass and
 momentum, while the second moment is uniformly bounded
 in time, conservation of both mass and momentum pass to the limit. We remark that
 these conservations are difficult to prove directly on the
 Fokker-Planck equation, due to the fact that $\phi(v) = v$ does
 not belong to $\bigF_s(\real)$.
 Hence we proved

\begin{theorem}\label{FPmain}
 Let the probability density $f_0 \in \bigM_0(\real)$,  and let the symmetric
random variable $Y$ which characterizes the kernel have a density
in $\bigM_{2+\alpha}$, with $\alpha > \delta$. Then, as $\gamma
\to 0$, $\sigma \to 0$ in such a way that $\sigma^2 =
\lambda\gamma$  the weak solution to the Boltzmann equation for
the scaled density $g_\gamma(v,\tau)=f(v,t)$, with $\tau = \gamma
t$ converges, up to extraction of a subsequence, to a probability
density $g(w,\tau) $. This density  is a weak solution of the
Fokker-Planck equation \rf{FP}, and it is such that the average
opinion is conserved.
\end{theorem}

\section{Other Fokker-Planck models of opinion formation}

Theorem \ref{FPmain} can be generalized in many ways. Always remaining with the
simplification of Remark \ref{max}, we can choose a general function $\Ci$ into
the interaction rule \fer{trade_rule}. The main difference with respect to the
proof of Theorem \ref{FPmain} is the evaluation of the first order term into
\fer{form}, which, using the fact that the mean value of $\Theta$ is zero,
reads
 \be \int_{\real^2}
 \Ci (w_* - w) \phi'(w)g(w) g(w_*) dw_*\,d w = \int_{\real}
 \Ci (m(\tau) - w) \phi'(w)g(w)\,d w,
  \ee
  where
 $m(\tau)$ is the value of the average opinion at time $\tau \ge 0$,
 \[
 m(\tau)=  \int_{\real}w g(w,\tau) d w.
 \]
 Note that, by \fer{resc}
 \be\label{meangrow}
m(\tau)=  \int_{\real}w g(w,\tau) d w =  \int_{\real}w f(w,t) d w.
 \ee
 Hence by \fer{grow5} the evolution of $m(\tau)$ obeys the law
   \be\label{grow55}
  \frac{d m(\tau)}{d\tau} = m(\tau)
 \int_{\real} \Ci g(w,\tau)\, d w  -
\int_{\real} w\Ci g(w,\tau)\, d w.
 \ee
Finally, in the limit $\g \to 0$ we obtain that $g(w,\tau)$
satisfies the Fokker-Planck equation
 \be\label{FP11}
 \frac{\partial g}{\partial \tau} = \frac \lambda 2\frac{\partial^2 }{\partial w^2}\left(\Di^2
 g\right) + \frac{\partial }{\partial w}\left(\Ci(w -m(t))
 g\right).
 \ee
\begin{remark}\label{meanv} {\rm The presence of a general propensity function
 $\Ci$ introduces a difficult to treat nonlinearity
into the Fokker-Planck equation. The nonlinearity is due to the
fact that the average opinion is no more constant, and  the
evaluation of the drift term requires the evaluation of
\fer{grow55}. }
\end{remark}

While to our knowledge the Fokker-Planck equation \fer{FP11} has never been
considered before, linked pure diffusion and drift equations have been recently
introduced in \cite{SL}. These equations, in our picture, refer to  diffusion
dominated ($\sigma^2/\gamma =
 \infty$) or compromise dominated ($\sigma^2/\gamma =
0$) limits. Looking at the proof of Theorem \fer{FPmain}, it is almost
immediate to conclude that the diffusion dominated limit takes into account
only the second-order term into the Taylor expansion. To verify this, suppose
that
 \[
 \frac{\s^2}{\g^\a} \to \lambda \, , \quad \a <1.
 \]
  Then  we can set
 \be\label{resc1}
 \tau = \gamma^\a t, \quad g(w,\tau) = f(w,t),
 \ee
 where now $g(w,\tau)$ satisfies
  \be
\frac{d}{d\tau}\int_\real g(w) \phi(w)\,dw  =
\frac1{\gamma^\a}\int_{\real^2}\int_{\Ball^2}\Theta(\eta)\Theta(\eta_*)
 [  (\gamma (w_* - w) + \eta \Di) \phi'(w)\nonumber +
  \ee
  \be\label{form1}
  + {1 \over 2}
(\gamma (w_* - w) + \eta \Di)^2 \phi''(w) ]
 g(w)g(w_*)  d w_*\,d w\,d \eta\,d\eta_*  + R(\gamma, \sigma),
 \ee
 with obvious meaning of the remainder.
 Since $\a <1$, the first order term in the Taylor expansion
 vanishes in the limit, and $g$ satisfies the diffusion equation
 \be\label{diff}
\frac{\partial g}{\partial \tau} = \frac \lambda 2\frac{\partial^2
}{\partial w^2}\left(\Di^2
 g\right).
 \ee
The choice
 \be
 \Di = \sqrt{ 1 -w^2}, \quad \lambda =2,
 \ee
brings to the diffusion equation
 \be
\label{diff1} \frac{\partial g}{\partial \tau} = \frac{\partial^2
}{\partial w^2}\left[(1-w^2)
 g\right].
 \ee
 This diffusion equation has been derived in a mean field approximation \cite{SL} to describe
 the evolution of the Sznajd model in Ochrombel simplification \cite{Och} on
 a complete graph of $N$ nodes in case of two opinions. The same equation has
 been shown to describe the former model in case of a large number
 $q$
 of opinions. The variable now represents the mean distribution of occupation
 numbers and the limit is taken as both $N$ and $q$ tend to
 infinity at the same rate.

 Likewise, the compromise dominated ($\sigma^2/\gamma =
0$) limit can be considered. In this case,
  \[
 \frac{\s^2}{\g^\a} \to \lambda \, , \quad \a >1,
 \]
  and  we can set
 \be\label{resc2}
 \tau = \gamma t, \quad g(w,\tau) = f(w,t).
 \ee
The diffusion part disappears in the limit and we obtain the pure
drift equation
 \be\label{drif}
 \frac{\partial g}{\partial \tau} =  \frac{\partial }{\partial w}\left(\Ci(w -m(t))
 g\right).
 \ee
 Note that, due to \fer{resc2} the evolution of the mean opinion
 $m(t)$ obeys the law \fer{meangrow}. The choice
  \[
  \Ci = { 1 -w^2}
  \]
   has been considered in \cite{SL}. In this case
 \be\label{drift}
 \frac{\partial g}{\partial \tau} =  \frac{\partial }{\partial w}\left((1-w^2)(w -m(t))
 g\right),
 \ee
 where
  \be
  \frac{d m(\tau)}{d\tau} =  - m(\tau)
 \int_{\real} w^2 g(w,\tau)\, d w  +
\int_{\real}w^3 g(w,\tau)\, d w.
 \ee
Thus, our equation differs from the pure drift in magnetization
obtained in \cite{SL} as the mean field limit of the Sznajd model
\cite{SWS} in case of two opinions. There the first-order partial
differential equation reads
 \be\label{drift1}
 \frac{\partial g}{\partial \tau} = - \frac{\partial }{\partial
 w}\left((1-w^2)w
 g\right).
 \ee
 Note  that the sign in front of the drift \fer{drift1} is now opposite
 to the sign in \fer{drift}, and the equation is now linear in
 $g$, even if the evolution of the mean opinion is not closed.

\section{Stationary solutions of the Fokker-Planck opinion model}

In this Section we analyze in some details various cases of the
interaction dynamics in the Boltzmann equation from which it is
possible to derive a Fokker-Planck equation with a explicitly
computable steady state. The structure of the steady state then
represents the formation of opinion consequent to the choice of
the interaction dynamics. In most cases, we are forced to suppose
$\Ci =1$, which implies conservation of the average opinion, and
Fokker-Planck \fer{FP} as underlying quasi-invariant opinion
limit. For any of these choices, we briefly discuss the link
between $\Di$ and the maximal support $\Ball$ of the diffusion
variable.

The first model of the diffusion  dependence on opinion we propose
is
 \[
 \Di = 1 - w^2.
 \]
 Since $\Ci=1$, the interaction rules are
 \[
w'   =  w - \gamma( w- w_*) + \eta (1-w^2)
 \]
 \[
 w_*'  =  w_* - \gamma (w_*- w) + \eta_* (1-w_*^2)
 \]
In this case  $ |\eta|(1+|w|) \le 1-\g$ implies $|w'| \le 1 $.
Hence, for a given opinion $\Ball =\left( -(1-\g)/(1+|w|),
(1-\g)/(1+|w|)\right)$, which shows that when the opinion $w$ is
close to zero and $\g$ is small, the effects of the global access
to information can move the opinion towards extremals. This
possibility reduces as soon as $|w|$ increases. The steady state
distribution of opinion is a solution to
 \be\label{FPstaz} \frac
\lambda 2\frac{\partial }{\partial w}\left((1-w^2)^2
 g\right) + (w -m)
 g= 0
 \ee
where $m$ is a given constant (the average initial opinion)
$-1<m<1$. The solution to \fer{FPstaz} is easily found as
 \be
g_\infty(w) = c_{m,\lambda} \left(
1+w\right)^{-2+m/(2\lambda)}\left(
1+wv\right)^{-2-m/(2\lambda)}\exp\left\{ -
\frac{1-mw}{\lambda(1-w^2)}\right\},
 \ee
where the constant $c_{m,\lambda}$ is such that the mass of
$g_\infty$ is equal to one. Note that the presence of the
exponential assures that $g_\infty(\pm 1) = 0$. The solution is
regular, but not symmetric unless $m=0$. Hence, the initial
opinion distribution reflects on the steady state through the mean
opinion. In any case, the stationary distribution has two picks
(on the right and on the left of zero) with intensity depending on
$\lambda$.

A similar result is expected from the choice
\[
 \Di = 1 - |w|.
 \]
As discussed in Section \ref{model} $ |\eta| \le 1-\g$ implies
$|w'| \le 1 $, and $\Ball =\left( -(1-\g), 1-\g\right)$. Once
more, the support of the random variable, for $\g$ small, covers
the whole domain of opinions.  The steady state distribution of
opinion is a solution to
 \be\label{FPstaz1} \frac
\lambda 2\frac{\partial }{\partial w}\left((1-|w|)^2
 g\right) + (w -m)
 g= 0
 \ee
where $m$ is a given constant (the average initial opinion)
$-1<m<1$. The solution to \fer{FPstaz} is easily found as
 \be
g_\infty(w) = c_{m,\lambda} \left( 1-|w|\right)^{-2-
2/\lambda}\exp\left\{ - \frac{1-mw/|w|}{2\lambda(1-|w|)}\right\},
 \ee
where, as usual, the constant $c_{m,\lambda}$ is such that the
mass of $g_\infty$ is equal to one. We remark that the low
regularity of $\Di$ reflects on the steady solution, which has a
jump in $w=0$. The jump disappears only when the mean $m=0$, and
only in this case we have a symmetric distribution. As in the
first case,  the presence of the exponential assures that
$g_\infty(\pm 1) = 0$, and the initial opinion distribution
reflects on the steady state through the mean opinion.

Last, we consider
\[
 \Di = \sqrt{1 - w^2}
 \]
in the Fokker-Planck model \fer{FP}. The steady state distribution
of opinion solves
 \be\label{FPstaz2} \frac
\lambda 2\frac{\partial }{\partial w}\left((1-w^2)
 g\right) + (w -m)
 g= 0,
 \ee
and equals
 \be g_\infty(w) = c_{m,\lambda} \left(\frac 1{
1 +w}\right)^{1 - (1+m)/\lambda}\left(\frac 1{ 1-w}\right)^{1 -
(1-m)/\lambda}.
 \ee
As usual, the constant $c_{m,\lambda}$ is such that the mass of
$g_\infty$ is equal to one. Since $-1 <m<1$, $g_\infty$ is
integrable on $\real$. Differently from the previous cases,
however $g_\infty (w)$ tends to infinity as $w \to \pm 1$, and it
has no peaks inside the interval $\real$. The explanation comes
out from a deep insight into the connection between $\Di$ and the
support $\Ball$ in this case.

It is immediate to verify that, in order to satisfy the constraint in
\fer{trade_rule}, one can not choose directly $\Di = \sqrt{1 -w^2}$. Within
this choice, in fact, choosing $w^* =1$, the first equality in \fer{trade_rule}
gives for $\eta$ the upper bound
 \[
\eta \le (1-\g)\frac{(1-w)}{\sqrt{1 -w^2}}.
 \]
Since the right-hand side converges to zero as $w \to 1$, it follows that there
is no way to satisfy the constraint. A different choice which gives in the
limit the Fokker--Planck equation with $\Di = \sqrt{1- w^2}$, is the following.
For any given $\g$, we set
 \[
 \Di = \sqrt{\left( 1 - (1+ \g^p)w^2 \right)_+}, \quad p >0,
 \]
where $f_+$ denotes as usual the positive part of $f$, that is
$f_+ = f$ if $ f>0$, while $f_+ = 0$ if $ f \le 0$. In this case,
one can show that it is sufficient for $\eta$ to satisfy the
condition
 \be
 |\eta| \le a_\g = \frac{1-\g}{\sqrt{1+\g^p}}\, \g^{p/2} ,
 \ee
 to respect the constraint on post-interaction opinions. To give
 an example, let us assume that $\Theta$ is uniformly distributed
 on the interval $-a_\g , a_\g$. Then $\sigma^2$ behaves like
 $\g^{(3p)/2}$, and it is enough to set $p = 2/3$ to obtain
 $\lambda =1$. The previous discussion shows that the
 choice $\Di = \sqrt{1-w^2}$ in the Fokker-Planck equation
 \fer{FP} corresponds to a kinetic interaction in which diffusion
 is of the order of $\g^p$, where $p$ is taken so that
 $\sigma^2/\g$ tends to a finite limit $\lambda$ as $\g \to 0$. In
 this case, the {\it smallness} of the interval of diffusion
 produces the peaks on $w = \pm 1$.

 A interesting feature of the Fokker-Planck equation
\be\label{FP111}
 \frac{\partial g}{\partial \tau} = \frac \lambda 2\frac{\partial^2 }{\partial
 w^2}\left( (1-w^2)
 g\right) + \frac{\partial }{\partial w}\left((w -m)
 g\right),
 \ee
 is that it leads to close evolution of moments. We remark that
equation \fer{FP111} is to be studied with the conservation of both mass and
momentum.

\section{Conclusions}

We introduced and discussed here some kinetic models of opinion formation based
on binary interactions involving both compromise and diffusion properties in
exchanges between individuals. A suitable scaling of compromise and diffusion
allows to derive Fokker-Planck equations for which it is easy to recover the
stationary distribution of opinion. Among these Fokker-Planck equations, one is
emerging \fer{FP111} and takes the role of the analogous one obtained in
\cite{BM, CPT} for the evolution of wealth. The main feature of this equation
is that moments can be evaluated in closed form. Further numerical studies are
in progress to understand the evolution of opinion density for various choices
of the underlying functions $\Ci$ and $\Di$. In particular, when $\Ci$ is not
linear, the evolution of moments is far from being completely understood.

\bigskip \noindent {\bf Acknowledgment:} The author acknowledges support from the IHP
network HYKE ``Hyperbolic and Kinetic Equations: Asymptotics,
Numerics, Applications'' HPRN-CT-2002-00282 funded by the EC., and
from the Italian MIUR, project ``Mathematical Problems of Kinetic
Theories''.

\end{document}